\documentclass[prl,aps,nofootinbib,showpacs,twocolumn,preprintnumbers,10pt]{revtex4}
\usepackage{amsmath,amssymb,graphicx,epsfig,hyperref,breakurl}

\begin{document}

\title{   CP Violation Effects on the Measurement of the
   CKM Angle  $\gamma$  from $B \to  DK$}
\author{ Wei Wang \footnote{Email:weiwang@hiskp.uni-bonn.de}
}
\affiliation{Helmholtz-Institut f{\"u}r Strahlen-und Kernphysik  and  Bethe Center for Theoretical Physics, Universit{\"a}t 
Bonn, D-53115 Bonn, Germany
}

\begin{abstract}
Inspired  by the unexpectedly large difference between the CP violation of $D$ decays into $K^+K^-$ and $\pi^+\pi^-$, we explore the impact on the extraction of $\gamma$ via the $B\to DK$ process with the $D$ meson reconstructed in the $K^+K^-,\pi^+\pi^-$ final state. 
We show that the extracted results for $\gamma$ can be shifted by ${\cal O}(A_{CP} / r_B^K)$, where $A_{CP}$ is the direct CP asymmetry in $D$ decays and $r_B^K$ is the ratio of the decay amplitudes of $B^-\to \bar D^0K^-$ and $B^-\to  D^0K^-$. Using the recent data on CP asymmetry, we demonstrate the correction to physical observables in $B\to DK$ can reach $6\%$, which  corresponds to the shift of $\gamma$ by roughly $5^\circ$.   The remanent corrections depend on the strong phase of the $D$ decays, but are less than $0.5^\circ$. With the increasing precision in the $\gamma$ determination on the LHCb experiment  and Super B factories, the inclusion of CP violation of $D$ decays will therefore soon become important. 
\end{abstract}

\pacs{13.25.Hw,12.15.Hh}
\maketitle
 

{\it Introduction.}---The fundamental role in the standard model (SM) description of CP violation is played by the unitary Cabibbo-Kobayashi-Maskawa (CKM) mixing matrix. 
The constraints on this matrix can be represented as triangles,
the lengths of whose sides are the moduli of  CKM matrix element products, while the angles
represent relative phases.
Among the three angles  $(\alpha,\beta,\gamma)$ of the   $(bd)$ unitarity triangle, satisfying  the constraint $\alpha+\beta+\gamma=180^\circ$,   $\gamma$ is least well known,  with a precision of roughly $10^\circ$. This is one of  the main sources of   current uncertainties on the apex of the unitary triangle~\cite{Asner:2010qj}.



In contrast with the $\alpha,\beta$ angles whose determination is  often challenged by the  loop penguin pollutions on theoretical side, one of the most intriguing properties of the angle $\gamma$ is that it can be measured in a way independent of hadronic uncertainty~\cite{Gronau:1991dp,Atwood:1996ci,Giri:2003ty}. That makes use of the tree-dominated processes $B\to DK$.  With a large amount of data accumulated in the future,  the LHCb  will be  able to diminish the errors in $\gamma$ to about $4^\circ$ from the tree-dominated processes $B\to DK$~\cite{Bediaga:2012py}, while on the SuperB factories the error is further reduced to $2^\circ$ \cite{Bona:2007qt,Aushev:2010bq}.  Hopefully with these results,  one may be able to  authenticate    the  unitarity of   CKM matrix, and thus the relevance of New Physics   for the  phenomena of  flavor physics or not.


Recently one of the most exciting measurements by   LHCb collaboration~\cite{Aaij:2011in}, confirmed by CDF~\cite{Collaboration:2012qw} and Belle~\cite{Ko:2012px} collaborations,  is the CP violation (CPV) in charm sector. 
These three collaborations have found nonzero  difference of  CP asymmetries (CPAs) which are much larger than the SM expectation. The averaged results are~\cite{Asner:2010qj}
\begin{eqnarray}\label{ACP_averaged}
\begin{split}
\Delta a_{CP}&=a_{CP}(K^+K^-)-a_{CP}(\pi^-\pi^+)\\
&= (-0.74 \pm 0.15 )\%.
\end{split}
\end{eqnarray} 
Here  $a_{CP}(f)$ is the time integrated CP asymmetry for $D$ decaying into a CP eigenstate:
\begin{eqnarray}
 a_{CP}(f)= \frac{\Gamma(D^0\to f)-\Gamma(\bar D^0\to f)}{\Gamma(D^0\to f)+\Gamma(\bar D^0\to f)}. 
\end{eqnarray}
The dominant  contribution to $\Delta a_{CP}$ in Eq.~\eqref{ACP_averaged} is from the direct CP violation of $D^0$ decays~\cite{Asner:2010qj}:
\begin{eqnarray}
 \Delta A^{\rm dir}_{CP} = (-0.678 \pm 0.147 )\%, \label{delta_ACP}
 \end{eqnarray}
 which is about $4.6\sigma$ away from 0. 


On the one hand it is   desirable to look for more precise measurement to establish the large CPA and identify possible future experimental tests able to distinguish the standard model vs new physics interpretations. On the other hand, it is of great importance to investigate the   impact of $a_{CP}$ on the known phenomena. 
In this work, we are interested in the effects of the nonzero CPA on the determination of the $\gamma$ angle, in particular  in the method to use the $B\to DK$ with $D$ decays into CP eigenstate $K^+K^-$ or $\pi^+\pi^-$, the so-called the GLW method~\cite{Gronau:1991dp}.    We will demonstrate that the CPA effects can shift the $\gamma$ by a few degrees. These  corrections are larger than or comparable  with the future experimental precision and thus  must be included in the future measurement. 
 
    
{\it  $\gamma$ measurement via $B\to D_{CP}K$.}---Let us  start with a review of the    approach  with the negligence of  the CP violation effects.    The GLW  method  uses the fact that the six decay amplitudes of $B^\pm \to (D^0, \bar D^0, D_{CP}^0)K^\pm$ form two triangles in the complex plane,  graphically representing the   identities
\begin{eqnarray}
 \sqrt 2 A(B^+\to D_\pm^0 K^+) = A(B^+ \to D^0K^+) \nonumber\\ \pm A(B^+\to \bar D^0K^+), \nonumber\\
 \sqrt 2 A(B^-\to D_\pm^0 K^-) = A(B^- \to D^0K^-) \nonumber\\ \pm A(B^-\to \bar D^0K^-),\label{eq:identity}
\end{eqnarray} 
where the convention $CP|D^0\rangle =|\bar D^0\rangle$ has been adopted and $D^0_+(D^0_-)$ is  the CP even (odd) eigenstate.  The   $K^+K^-$ and $\pi^+\pi^-$ final states will mainly project out the  $D^0_+$, which will be considered in the following. 
 Measurements of the six decay rates  will  completely determine the sides and apexes of the two triangles, 
 in particular the relative phase between $A(B^- \to \bar D^0K^-)$ and $A(B^+ \to D^0K^+)$ is $2\gamma$. 
Since the identities in Eq.~\eqref{eq:identity} holds irrespective of the strong phase in the decay, this method is free of hadronic uncertainties and  is   believed  to be theoretically clean.

Different with the loop-induced processes, tree-dominated processes are unlikely affected by the new physics degrees of freedoms and thus the measurement of $\gamma$ provides a benchmark of extraction of the CKM parameters. 

The shape of the two triangles is governed by  two quantities  
\begin{eqnarray}
 r_{B}^K \equiv\left|{A(B^-\to \bar D^0 {K^-})}/{A(B^-\to D^0 K^{-})}\right|,\nonumber\\
 \delta_{B}^{K} \equiv arg\left[{e^{i\gamma} A(B^-\to \bar D^0 K^{-})}/{A(B^-\to D^0 K^{-})}\right],\nonumber
\end{eqnarray}
with the world averages for these parameters~\cite{CKMfitter}
\begin{eqnarray}
 r_B^K= 0.107\pm 0.010,\;\;
 \delta_B^K=( 112^{+12}_{-13})^\circ. 
\end{eqnarray}
The smallness of $r_B^K$ implies the mild sensitivity to $\gamma$ and introduces experimental difficulty. Thus 
  additional  methods using processes in which the $D^0$ and $\bar D^0$ meson can be accessed in the same final states are proposed~\cite{Atwood:1996ci,Giri:2003ty}.

{\it CPA effects.}---We now study the effects of  CP violation but neglect the small D mixing effects~\cite{Grossman:2005rp}. 
In particular, we ask how large is the  error   introduced in the extracted value when the analysis is done assuming
no CPV.   

 The  $D$ meson decay amplitudes can be generically decomposed as
 \begin{eqnarray}
 A(D^0\to f) &=& T_{D}^f (1+ r_D^f e^{-i\gamma + i \delta_D^f}),\nonumber\\
 A(\bar D^0\to f) &=& T_{D}^f (1+ r_D^f e^{i\gamma + i \delta_D^f}),
\end{eqnarray}   
where  $r_D^f$ is the ratio of the penguin and tree amplitudes in $D\to f$ decays, with $f=K^+K^-,\pi^+\pi^-$.  $\gamma$ and $\delta_D^f$ is the weak phase difference and strong phase difference respectively. 
The direct CP asymmetry is predicted as
\begin{eqnarray}
\label{eq:DirCPA}
&&A_{CP}^{dir}(D^0\to f) = \frac{2r_D^f \sin\gamma \sin\delta_{D}^f}{1+ (r_D^{f})^2 +2r_D^f \cos\gamma \cos \delta_{D}^f}.
\end{eqnarray}
The  data on $A_{CP}$ in Eq.~\eqref{delta_ACP} indicates $|r_D^f| \sim {\cal O}(10^{-3})$.



Including the CP violating amplitudes of the $D$ decays, we arrive at  
\begin{widetext}
\begin{eqnarray}
 \sqrt{2} A(B^-\to D_{+}^0  (\to f)  K^-)  &=& A(B^-\to D^0 K^-)  T_D^{f}\big[(1+ r_D^f e^{-i\gamma + i \delta_D^f})+ r_{B}^K e^{-i\gamma +i\delta_B^K}  (1+ r_D^f e^{i\gamma + i \delta_D^f})\big],\nonumber\\
 \sqrt{2} A(B^+\to D_{+}^0  (\to f) K^+)  &=& A(B^+\to D^0 K^+)  T_D^{f}\big[r_{B}^K e^{i\gamma +i\delta_B^K} (1+ r_D^f e^{-i\gamma + i \delta_D^f})+  (1+ r_D^f e^{i\gamma + i \delta_D^f})\big],
\end{eqnarray}    
 \end{widetext}
which apparently will spoil the identities in Eq.~\eqref{eq:identity} since it is not possible to simultaneously  eliminate the nontrivial dependence on $r_D^f$ and $\delta_D^f$ in the two amplitudes.  

The physical observables to be experimentally measured and used to extract the CKM angle $\gamma$  are given as
\begin{eqnarray}
 R_{+}^K &=&2\frac{{\cal B}(B^-\to D^0_{+} K^-)+{\cal B}(B^+\to D^0_{+} K^+)  }{{\cal B}(B^-\to D^0K^-) +{\cal B}(B^+\to \bar D^0 K^+) }\nonumber\\
&=&1+(r_{B}^K)^2  \nonumber\\
&& + \frac{2r_B^K \cos\delta_B[ (1+(r_D^f)^2)\cos\gamma+ 2 r_D^f   \cos\delta_D^f]}{1+ (r_D^{f})^2 +2r_D^f \cos\gamma \cos \delta_{D}^f}, \nonumber\\
&\equiv & 1+(r_{B}^K )^2+ 2r_{B}^K  \cos\delta_{B}^K \cos\gamma_{eff},\label{eq:RplusK_corrections}\end{eqnarray}
\begin{eqnarray}
 A_{+}^K &=&\frac{{\cal B}(B^-\to D^0_{+} K^-)-{\cal B}(B^+\to D^0_{+} K^+)  }{{\cal B}(B^-\to D^0_{+} K^-) +{\cal B}(B^+\to D^0_{+} K^+) }\nonumber\\
&=& \frac{1}{ R_{+}^K }  \bigg[(1-(r_B^K)^2)  A_{CP}^{dir} (D^0\to f)  \nonumber\\
&& +  \frac{ 2r_B^K  (1+(r_D^f)^2) \sin\delta_{B}^K \sin\gamma }{ 1+(r_D^f)^2 +2r_D^f \cos\delta_D^f \cos\gamma}\bigg] \nonumber\\
&\equiv &  2r_B^K \sin\delta_{B}^K \sin\gamma_{eff} /R_{+}^K, \label{eq:AplusK_corrections}
\end{eqnarray}
where the last lines in the above equations correspond to the expressions with no CPV effects. 
In the above expressions, we have substituted  the CP averaged branching ratio of $D\to f$ in the processes involving  $D^0_+$. 
These two  equations explicitly show the CPA effects on the experimental observables and are one of the main findings of this work.

{\it Results.}---An interesting observation is that the $ A_{+}^K $ in Eq.~\eqref{eq:AplusK_corrections} receives new contributions proportional to the direct CPA in $D$ decays. 
Neglecting terms suppressed by ${\cal O}(r_D^f)$, the dominant  correction to  $\sin\gamma$  is proportional to $A_{CP}^{dir} (D^0\to f)/(2r_B^K \sin\delta_{B})$.  
One consequence is that  the value for $\gamma$ obtained from $K^+K^-$ final state and the one from  $\pi^+\pi^-$ final states will  differ roughly by  $5^\circ$.  These effects can still be incorporated without any hadronic uncertainty once the data on the direct CPA is available. 

For illustration, we consider three general pattens for the CPV: (i)   $A_{D}^{K^+K^-}= - A_{D}^{\pi^+\pi^-}= \Delta A_{CP}/2$; (ii) $A_{D}^{K^+K^-}=  \Delta A_{CP}$ and $ A_{D}^{\pi^+\pi^-}=0$; (iii) $A_{D}^{K^+K^-}= 0$ and $ A_{D}^{\pi^+\pi^-}=- \Delta A_{CP}$. 
In the first patten, considering the dominant corrections in the  expressions for $A_{+}^K$ in Eq.~\eqref{eq:AplusK_corrections} to extract the $\gamma$, we find that the  difference $\Delta \gamma= \gamma_{eff}-\gamma$ is  roughly $-2.5^\circ$ in the $K^+K^-$ final state, while it is $2.5^\circ$ in the $\pi^+\pi^-$ final state.   In the second patten of CPV, the modification in $K^+K^-$ is roughly $-5^\circ$ while the extracted value for $\gamma$ from the  $\pi^+\pi^-$ channel is almost unchanged. The last patten is analogous to the second one except that the shift in $\pi^+\pi^-$ is $5^\circ$ while the change in  $K^+K^-$ mode is negligible.  The errors from the current LHCb measurements on $A_{+}^K$ from the $K^+K^-$ and $\pi^+\pi^-$ final states are too large to give any conclusive result~\cite{LHCb-CONF-2012-032}
\begin{eqnarray}
 A_{+}^K (K^+K^-) = 0.148\pm 0.037\pm 0.010,\nonumber\\
 A_{+}^K (\pi^+\pi^-) = 0.135\pm 0.066\pm 0.010. 
\end{eqnarray}

\begin{figure}\begin{center}
\includegraphics[scale=0.5]{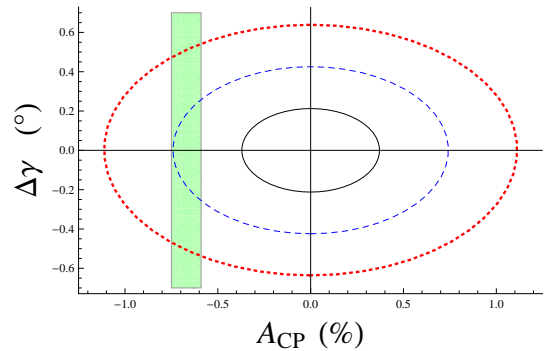}
\caption{Effect of CP violation  on the extraction of   $\gamma$ via the $R_{+}^K $. The sold (black), dashed(blue), and dotted (red) lines correspond to $r_{D}^f$=0.002, 0.004 and  0.006 respectively.  The shadowed region is the CPA for $D^0\to K^+K^-$ from the experimental data:  $A_{CP}^{K^+K^-}= (-0.34\pm 0.07)\%$, where we have assumed the U-spin symmetry for the CP asymmetry $A_{CP}^{K^+K^-}= -A_{CP}^{\pi^+\pi^-}= \Delta A_{CP}/2$.   } \label{fig:gamma_R_CP}
\end{center}
\end{figure}

Apart from the corrections proportional to the direct CPA,  there are still more suppressed  terms depending on the strong phase $\delta_{D}^f$ as shown in Eq.~\eqref{eq:RplusK_corrections} and Eq.~\eqref{eq:AplusK_corrections}. 
The suppressed  CP violation effect in  the $R_{+}^K $  is shown in Fig.~\ref{fig:gamma_R_CP}. The solid (black), dashed(blue), and dotted (red) lines correspond to $r_{D}^f$=0.002, 0.004 and  0.006 respectively.  The shadowed region is the data on the  CPA for $D^0\to K^+K^-$ in the first patten for CPV:  $A_{CP}^{K^+K^-}= (-0.34\pm 0.07)\%$. From the figure we can see the difference between the results for  $\gamma$ with and without the CPV effects is up to $0.5^\circ$, comparable with the experimental  accuracy ($0.8^\circ$) in $\gamma$ after the LHCb upgrade~\cite{Bediaga:2012py}.


{\it Discussions.}---The dominant corrections proportional to the  $A_{CP}^{\rm dir}$ can  be incorporated without any hadronic uncertainty, once the data on the direct CPA is available.   However  using only $B\to DK \to (K^+K^-, \pi^+\pi^-)$, it is unlikely  to eliminate the dependence on the strong phase for the rest terms in  Eq.~\eqref{eq:RplusK_corrections} and Eq.~\eqref{eq:AplusK_corrections}.  It might be possible when    results from the charm factory are available  and more   relevant channels like $B\to DK^*$ are used.

Including the new Belle measurement, the combined result for $\Delta A_{CP}$ is about $4.6\sigma$ away from 0 as shown in Eq.~\eqref{delta_ACP}.  Still to date the nonzero $\Delta A_{CP}$ is not well-established.    
With the LHCb result   based on a small fraction of data recorded so far, significant improvements   are expected in the near future. 
Moreover to explicitly  explore the CPV effects on the extraction of $\gamma$,  one also has to wait for our experimental colleagues to measure the CPV for $D^0\to K^+K^-$ and $D^0\to \pi^+\pi^-$ individually.  The present results for the individual CP asymmetries from CDF~\cite{Aaltonen:2011se} and Belle~\cite{Ko:2012px} collaborations  are consistent with zero and are not conclusive. 

On the experimental side, one may instead use the averaged results from the $K^+K^-$ and $\pi^+\pi^-$ final states. Such procedure will indeed smear most of the CPV effects in the first patten for CP violations, but not in  the other two patterns.

The analysis is similar  in channels like $B\to DK^*$ in which the $D$ meson is reconstructed in $K^+K^-,\pi^+\pi^-$ final states. 
In the  processes of $B\to DK^{*}_{0,2}$~\cite{Wang:2011zw},  which  are likely to  have  a larger ratio $r_{B}^{K^*_0}$ due to  comparable size  of the color-allowed and color-suppressed amplitudes, the CP violating corrections, proportional to $A_{CP}^{\rm dir}/r_B^{K^*0}$,    will be somewhat smaller.

At last, 
it is  worthwhile  to stress that the CP asymmetry may not affect the ADS method and the Dalitz plot method. In the former, the doubly-Cabibbo suppressed  (DCS) $D$  decays is used to increase the sensitivity while in the Dalitz plot method, the three-body $D\to K_S\pi^+\pi^-$ and $D\to K_SK^+K^-$ decays are used to reconstruct the $D_+$. The involved  Cabibbo allowed and DCS decay modes are typically having very small CPAs and thus the extraction of $\gamma$ is almost unaffected when neglecting the CPA.  Moreover one of the greatest advantages of the  Dalitz plot method is that   the CPA effects and the small DCS contributions can be explicitly incorporated when  enough  data is accumulated. 

 
 {\it Conclusions.}---After roughly 40 years since the proposal,  the CKM mechanism continues to provide a consistent description  of almost all available data on the flavor observables and CP violation with an impressive accuracy.  This great success implies  that the New Physics effects should be small,  and renders the precision predictions for the involved quantities particularly important.  Thus  the precise determination of the CKM angles is of great importance, for which   one of the most desirable objectives  in the ongoing and forthcoming experiments  is to reduce the errors in the involved entries of the unitary triangle. 

  What has been explored in this work is to  study the CPV effects in the GLW method for the extraction of CKM angle $\gamma$ and  showed  that the CPV corrections are of the order $ A_{CP} / r_B^K$. 
If CP violation in $D$ decays is ignored in the extraction of $\gamma$,  we have found the corrections can reach  $6\%$  and the resulting shift  of $\gamma$ is about $5^\circ$,  larger than or comparable with the precision  to be achieved  on the future LHCb ($4^\circ$) and SuperB ($2^\circ$) experiments.  With the increasing precision in the $\gamma$ determination by these experiments  the inclusion of CPV of $D$ decays will therefore soon become mandatory.  

{\it Acknowledgement.}---The author is grateful to  Chuan-Hung Chen, Malcolm John and Yue-Hong Xie for  useful discussions.  He also thanks Institute of High Energy Physics and Tianjin University for their hospitalities during his visit when part of this work was done.  This work is supported in
part by the DFG and the NSFC through funds provided to
the Sino-German CRC 110 ``Symmetries and the Emergence
of Structure in QCD".


\end{document}